\documentclass[12pt]{article}
\usepackage{graphics,epsfig}
\usepackage[english]{babel}
\newcommand{\cinst}[2]{$^{\mathrm{#1}}$~#2\par}
\newcommand{\crefi}[1]{$^{\mathrm{#1}}$}

\begin{document}


\thispagestyle{empty}



\begin{center}


 \vglue 1.0cm {\Large \textbf{A study of the neutrinoproduction of
 $\Phi$ and $D_s^+ $ mesons}}

\end{center}

\vspace{1.cm}

\begin{center}
{\large SKAT Collaboration}

 N.M.~Agababyan\crefi{1}, V.V.~Ammosov\crefi{2},
 M.~Atayan\crefi{3},\\
 N.~Grigoryan\crefi{3}, H.~Gulkanyan\crefi{3},
 A.A.~Ivanilov\crefi{2},\\ Zh.~Karamyan\crefi{3},
V.A.~Korotkov\crefi{2}

\setlength{\parskip}{0mm}
\small

\vspace{1.cm} \cinst{1}{Joint Institute for Nuclear Research,
Dubna, Russia} \cinst{2}{Institute for High Energy Physics,
Protvino, Russia} \cinst{3}{Yerevan Physics Institute, Armenia}
\end{center}
\vspace{100mm}

{\centerline{\bf YEREVAN  2009}}


\newpage
\vspace{1.cm}
\begin{abstract}

\noindent The neutrinoproduction of $\phi$ and $D^+_s$ mesons is
studied, using the data obtained with the SKAT bubble chamber at
the Serpukhov accelerator. It is found that the $\phi$ production
occurs predominantly in the forward hemisphere of the hadronic
c.m.s. (at $x_F > 0$, $x_F$ being the Feynman variable), with the
mean yield strongly exceeding the expected yield of directly
produced $\phi$ mesons and varying from $\langle n_{\Phi}(x_F > 0)
\rangle = (0.92\pm0.34) \cdot 10^{-2}$ at $W > $ 2 GeV up to
$(1.23\pm0.53) \cdot 10^{-2}$ at $W >$ 2.6 GeV and $(1.44\pm0.69)
\cdot 10^{-2}$ at $W >$ 2.9 GeV, $W$ being the invariant mass of
the hadronic system. The yield of leading $D^+_s$ mesons carrying
more than $z = 0.85$ of the current $c$- quark energy is
estimated: at $W > 2.9$ GeV, $\langle n_{D^+_s}(z > 0.85) \rangle
= (6.64\pm1.91) \cdot 10^{-2}$. It is shown, that the shape of the
$\phi$ meson differential spectrum on $x_F$ is reproduced by that
expected from the $D^+_s \rightarrow \phi X$ decays which,
however, can account for only the half of the measured $\phi$
yield.

\end{abstract}

\newpage
\setcounter{page}{1}
\section{Introduction}

\noindent The total and differential yields of hadrons in
leptonuclear reactions reflect the space-time structure of the
quark string fragmentation and the formation of hadrons, both
produced directly or as a result of secondary intranuclear
interactions or originated from the decays of resonances. The
latters play a significant role in the production of stable
hadrons (pions and kaons). For example, in the case of
neutrino-induced reactions the fraction of pions from the decays
of light meson resonances (up to masses $\sim$ 1 GeV/$c^2$)
composes about 20-40\%, depending on the pion type and the
neutrino energy $E_\nu$ (see \cite{ref1} and references therein).
The information on the space-time structure of the hadron
formation based only on the data on stable hadrons would,
therefore, be rather incomplete and should be complemented by the
data on the resonance production. The latter data themselves
provide more direct information on the quark string fragmentation,
since they are, in general, expected to be less contaminated by a
contribution from the decays of the higher-mass resonances. This
contribution can be rather different for various resonances and in
various kinematical regions of their production. In particular, in
neutrino-induced reactions the production of a favorable
resonance, which can carry a large fraction of the current quark
energy and hence occupies the forward hemisphere in the hadronic
c.m.s. (with $x_F > 0$, $x_F$ being the Feynman variable), is
contributed by the decay of the higher-mass resonances mainly at
$x_F < 0$, but to a much smaller extent at $x_F > 0$. The
situation with unfavorable resonances (not containing the current
quark) can be rather different.
An interesting example is $\Phi$ meson. \\
Note first of all, that the neutrinoproduction of direct $\Phi$
mesons, both unfavorable and especially favorable, is expected to
be strongly suppressed as compared to other favorable vector
mesons, for example, $\omega$ meson (with the same spin and
isospin) which can be produced via the main subprocess $\nu
d\rightarrow\mu^ -u$ followed by the recombination of the current
$u$- quark with a $\bar{u}$- antiquark created at the string
breaking. The production of a favorable $\Phi$ meson, which
proceeds via the subprocess $\nu \bar{u}\rightarrow \mu^- \bar{s}$
followed by the recombination with a $s$- quark, is
triple-suppressed as compared to the $\omega$ production due to:
i) the suppressed magnitude of the $V_{us}$ element of
Cabibo-Kobayashi-Maskawa matrix \cite{ref2}; ii) the strangeness
suppression in the quark string fragmentation \cite{ref3}; iii)
the smallness of the fraction of the nucleon momentum (integrated
over the Bjorken $x_B$ variable) carried by the $\bar{u}$-
antiquark as compared to that for the $d$- quark (see e.g.
\cite{ref4}). The product of these suppression factors is about $5
\cdot 10^{-4}$. Even when taking into account the contribution
from the competing recombination processes, e.g. $u \bar{u}
\rightarrow \pi^0, \rho^0, \eta, \eta^{'}$ for the case of
$\omega$ and $s \bar{s} \rightarrow \eta, \eta^{'}$ for the case
of $\phi$, the expected ratio of the yields of favorable $\phi$
and $\omega$ mesons remains very small, no more than $ 10^{-3}$.
\\ This value is in strict contrast with experimentally estimated
ratio $\langle n_{\Phi}(x_F > 0) \rangle /\langle n_{\omega}(x_F >
0) \rangle = 0.18\pm0.08$ at $x_F > 0$ \cite{ref1}, indicating on
the dominant role of indirect processes in the $\Phi$
neutrinoproduction. One of them is, probably, related to the
production of the charmed, strange $D^+_s$ meson via the
subprocess on a strange sea quark, $\nu s\rightarrow \mu^- c$,
followed by the recombination of the charm quark with the sea
$\bar{s}$- antiquark from the nucleon remnant, $(c
\bar{s})\rightarrow D^+_s$. Since here no creation of an extra $(q
\bar{q})$ pair is necessary, the $D^+_s$ meson can carry the
overwhelming fraction $z$ of the current $c$- quark energy and
then transfer a significant part of the latter to the $\Phi$ meson
in $D^+_s\rightarrow \Phi X$ decays (which occur with a summary
rate 16.1$\pm$1.6\% \cite{ref2}). An alternative subprocess,
leading to the $D^+_s$ production (with approximately the same
probability as in the previous case), is $\nu d \rightarrow \mu^-
c$ followed by the recombination of the $c$- quark with a
$\bar{s}$- antiquark created at the string breaking. In this case,
however, $D^+_s$ carries on an average a lesser energy fraction
$z$. \\
This work is devoted to the experimental study of the
neutrinoproduction of $\Phi$ and $D^+_s$ mesons, with a particular
aim to check the aforesaid mechanism of the indirect $\Phi$
production. In Section 2, the experimental procedure is described.
Section 3 presents the experimental data on the total and
differential yields of $\Phi$ meson. In Section 4, several decay
modes of $D^+_s$ meson are analyzed and an estimation of its
yields is inferred for the case of the leading $D^+_s$ meson
carrying the overwhelming fraction $(z > 0.85)$ of the current
quark energy. The results are summarized in Section 5.

\section{Experimental procedure}

\noindent The experiment was performed with SKAT bubble chamber
\cite{ref5}, exposed to a wideband neutrino beam obtained with a
70 GeV primary protons from the Serpukhov accelerator. The chamber
was filled with a propane-freon mixture containing 87 vol\%
propane ($C_3H_8$) and 13 vol\% freon ($CF_3Br$) with the
percentage of nuclei H:C:F:Br = 67.9:26.8:4.0:1.3 \%. A 20 kG
uniform magnetic field was provided within the operating chamber
volume. \\
Charged current ($CC$) interactions containing a negative muon
with momentum $p_{\mu} >$0.5 GeV/c were selected. Other negatively
charged particles were considered to be $\pi^-$ mesons, except for
the cases explained below. Protons with momentum below 0.6 GeV$/c$
and a fraction of protons with momentum 0.6-0.85 GeV$/c$ were
identified by their stopping in the chamber. Non-identified
positively charged particles were considered to be ${\pi}^+$
mesons, except for the cases explained below. Events in which
errors in measuring the momenta of all charged secondaries and
photons were less than 60\% and 100\%, respectively, were
selected. The mean relative error $\langle \Delta p/p \rangle$ in
the momentum measurement for muons, pions and gammas was,
respectively, 3\%, 6.5\% and 19\%. Each event is given a weight
which corrects for the fraction of events excluded due to
improperly reconstruction. More details concerning the
experimental procedure, in particular, the estimation of the
neutrino energy $E_{\nu}$ and the reconstuction of
$\pi^0\rightarrow 2\gamma$ and neutral strange particle decays can
be found in our previous publications \cite{ref6,ref7,ref8}. \\
The events with $3< E_{\nu} <$ 30 GeV were accepted, provided that
the reconstructed mass $W$ of the hadronic system exceeds
$W_{min}$ = 2 GeV. No restriction was imposed on the transfer
momentum squared $Q^2$. The number of accepted events was 4577
(5784 weighted events). The mean values of the kinematical
variables were $\langle E_{\nu} \rangle$ = 10.7 GeV, $\langle W
\rangle$ = 3.0 GeV, $\langle W^2
\rangle$ = 9.6 GeV$^2$, $\langle Q^2 \rangle$ = 2.8 (GeV/c)$^2$. \\
About 8\% of neutrino interactions occur on free hydrogen. This
contribution was subtracted using the method described in
\cite{ref9,ref10}. \\
When considering the production of resonances decaying into
charged kaons, the $K^-$ and $K^+$ hypothesis was applied,
respectively, for negatively charged particles and non-identified
positively charged particles with momenta $p > p_{cut} = 0.55$
GeV$/c$, introducing thereat proper corrections for the momentum
of these particles. It has been checked that the choice of lower
values of $p_{cut}$ does no practically influence the results
presented in next sections.

\section{The total and differential yields of $\phi$ meson}

The $K^+K^-$ effective mass distributions for three different
ranges of $x_F$ are plotted in Figure 1. The main problem in the
$\phi$ signal separation is the large background from
misidentified $\pi^+ \pi^-$ pairs having small effective masses.
The background related to the correlated low-mass $\pi^+ \pi^-$
from $\eta \rightarrow \pi^+ \pi^- \pi^0$, $\omega \rightarrow
\pi^+ \pi^- \pi^0$ and $\rho \rightarrow \pi^+ \pi^-$ decays was
subtracted from experimental distributions using the data on these
resonances obtained in \cite{ref1}. To describe the
remaining (combinatorial) background, two methods were applied: \\
i)The mixed event method, in which the shape of the background
distribution was determined combining kaons from different events
of the same topology and ranges of global kinematical variables
($E_\nu$, $W$). The normalization of the background distribution
was kept as a free parameter when fitting the experimental mass
distribution; \\
ii) The fitted background method, in which the background was
approximated by
\begin{equation}
BG(m) = B \cdot (m - 2m_K)^\beta \, \cdot \exp{(-\gamma m)} \, ,
\end{equation}
\noindent where $m_K$ is the charged kaon mass, while $B$, $\beta$
and $\gamma$ are free parameters (in some cases $\gamma$ was fixed
to zero). \\
The 'signal' distribution was parametrized by a Gaussian form
\begin{equation}
G_{\phi}(m) = \alpha_{\phi} \cdot \exp[{-\frac{{(m-m_{\phi})}^2}{2
\sigma^2_{\phi}}}] \, \, ,
\end{equation}
\noindent with the fixed pole mass $m_{\phi}$ = 1019 MeV and the
experimental resolution $\sigma_{\phi}$ = 4 MeV (estimated by
simulations, resulting in a FWHM value $\Gamma^{res}_{\phi}
\approx$ 10 MeV exceeding largely the $\phi$ natural width
$\Gamma^0_{\phi} \approx$ 4.3 MeV). \\
The mass distributions shown in Figure 1 were fitted as a sum of
the 'signal' and background distributions. The fit results for
both methods i) and ii) turned out to be practically the same.
 As it is seen, the data exhibit a clear $\phi$ signal at $x_F
> 0$, with the corresponding yield $\langle n_{\Phi}(x_F > 0)
\rangle = (0.92\pm0.34)\cdot 10^{-2}$ (corrected for unobserved
$\phi$ decay modes). No $\phi$ production is observed in the
backward hemisphere ($x_F < 0$), where the data are consistent
with the background distribution. As a result, the $\phi$ signal
in the full $x_F$- range turns out to be less expressed as
compared to that at $x_F > 0$, leading to a less accurate
estimation of the
total yield: $\langle n_{\Phi} \rangle = (1.19\pm0.61)\cdot 10^{-2}$. \\
We have also obtained the $W_{min}$- dependence of $\langle
n_{\Phi}(x_F > 0) \rangle$ and found it to increase continuously
from $\langle n_{\Phi}(x_F > 0) \rangle = (0.8\pm0.3) \cdot
10^{-2}$ at $W_{min}$ = 1.8 GeV \cite{ref1} to $\langle
n_{\Phi}(x_F > 0) \rangle = (1.23\pm0.53) \cdot 10^{-2}$ and
$(1.44\pm0.69) \cdot
10^{-2}$ at $W_{min}$ = 2.6 GeV and 2.9 GeV, respectively. \\
The differential spectrum $dn_{\phi}/dx_F$, compared to that for
$\rho^0$ meson \cite{ref11} (at $W > 2$ GeV), is presented in
Figure 2, from which two observations emerge. Firstly, unlike
$\Phi$ meson, the yield of $\rho^0$ in the backward hemisphere is
not much smaller than that in the forward hemisphere. This fact
can be explained \cite{ref11} by a contribution from secondary
intranuclear interactions of pions, $\pi N \rightarrow \rho^0 X$,
resulting in a $\rho^0$ multiplicity gain, mainly at $x_F < 0$.
Nothing similar happens with $\phi$, due to a much smaller
probability of the $\phi$ production in secondary interactions
(see Section 5 for details). Secondly, the yield of $\rho^0$ is
rather prominent at large $x_F
> 0.6$ $-$ a characteristic feature for the direct leptoproduction
of a favorable hadron. On the contrary, the yield of $\phi$ is
more concentrated around moderate positive values of $x_F$,
indicating on its origin from the decay of a favorable leading
resonance. A probably candidate for the latter is, as it was
mentioned in Introduction, the charmed, strange $D^+_s$ meson the
production of which is considered in the next section.

\section{The yield of $D^+_s$ meson}

As it was mentioned above, the $x_F$- dependence of the $\phi$
meson yield indicates on its origin from the decay of leading
$D^+_s$ mesons which can be produced, for example, in exclusive
reactions
\begin{equation}
\nu N \rightarrow \mu^- + N^* + D^+_s \, (or \, D^{*+}_s)  \, ,
\end{equation}
\noindent and
\begin{equation}
\nu N \rightarrow \mu^- + Y + D^+_s \, (or \, D^{*+}_s)  \, ,
\end{equation}
\noindent with the vector $D^{*+}_s$ state decaying into $D^+_s
\gamma$ (94.2\%) or $D^+_s \pi^0$ (5.8\%), a nucleon remnant $N^*$
(turning into a nucleon or a low-state resonance), and a hyperon $Y$. \\
In reactions (4) and (5), most of $D^+_s$ mesons carry the
overwhelming fraction $z$ of the current $c$- quark energy. The
mean yield of the leading $D^*_s$ vector meson (with $z > 0.75$)
was measured earlier in $\bar{\nu} Ne$- interactions, resulting in
$\langle n_{D_s^{*-}}{(z > 0.75)}\rangle = 0.051\pm0.016$
\cite{ref12}. One can assume, that the mean yield of the
pseudoscalar $D^-_s$ (direct production) composes 1/3 of that for
$D^{*-}$, i.e. the summary yield of $D^-_s$ is expected to be
about $0.068\pm0.021$, leading to the mean yield of the decay
$\phi$ mesons $\langle n_{\phi}^{dec}
\rangle = 0.014\pm0.006$ per $CC$ $\bar{\nu}Ne$- interaction. \\
Below we consider several hadronic decay modes of $D^+_s$ with
branching ratios exceeding a few percent (see Table). A rather
severe cut was applied on $z$ ($>$ 0.85) in order to reduce the
contribution from the background processes. The events with $W <$
2.9 GeV, i.e. below the threshold of the $D^+_s$ production, were
excluded (it should be, however, noted, that the experimental
resolution of $W$ is about 10\% in this experiment, and hence the
events with the estimated value of $W$ in between $2.6 < W < 2.9$
GeV can also have a small contribution to the $D^+_s$ production).
The experimental mass resolutions (estimated by simulations) for
considered channels are quoted in Table. \\

Table. The considered $D^+_s$ decay modes, experimental mass
resolutions, estimated yields at $z > 0.85$ and decay fractions.

\begin{center}
\begin{tabular}{|l|c|c|c|}
  \hline

&&&   \\
Decay mode&Mass resolution (MeV)&Estimated yield
&Decay fraction \\
&&(in $10^{-2}$)& (in \%) \cite{ref2} \\ &&& \\ \hline

a)$K^+K^- \pi^+$&22&0.56$\pm0.26$&~~5.50$\pm0.28$ \\
b) $K^+K^-\pi^+ \pi^0$&45&0.50$\pm$0.25&5.6$\pm$0.5 \\
c) $3\pi^+2\pi^-\pi^0$&42& 0.17$\pm$0.12&
4.9$\pm$3.2 \\
d) $\rho^+\eta', \eta'\rightarrow \rho^0 \gamma$&52 &0.25$\pm$0.20&3.6$\pm$0.6 \\
e) $\rho^+\eta, \eta\rightarrow \pi^+\pi^-\pi^0$&120 & 0.09$\pm$0.09 & 3.0$\pm$0.5 \\
f) $\rho^+\eta, \eta\rightarrow \gamma \gamma$&200 & 0.27$\pm$0.20
& 5.1$\pm$0.8
\\ \hline
\bf{sum}&&\bf{1.84$\pm$0.48} & \bf{27.7$\pm$3.4} \\
\hline
\end{tabular}
\end{center}

The main channels with a $K^+ K^-$ pair in the final states are:
a) $K^+ K^- \pi^+$ and b) $K^+ K^- \pi^+ \pi^0$. The contamination
from $\pi^+ \pi^-$ pairs is expected \cite{ref12} to be more
prominent at large (in the absolute value) negative values of
$\cos \vartheta^*_{KK}$, $\vartheta^*_{KK}$ being the angle
between the $K^+K^-$ direction in the $D^+_s$ candidate rest frame
and the direction of the Lorentz boost from the lab system to the
$D^+_s$ rest frame. We excluded the combination with $\cos
\vartheta^*_{KK} < -0.9$ where a significant enhancement of the
number of combinations was observed. For channel b) containing a
$\pi^0$, a cut 115 $< m_{\gamma\gamma} <$ 155 MeV was applied for
the $\gamma\gamma$ effective mass (the same cut was also applied
for channels c)$-$f) containing one or two neutral pions in the
final state). The effective mass distributions for channels a) and
b) are plotted in Figure 3. For both channels, an excess of events
around the $D^+_s$ mass is observed above the combinatorial
background which can be satisfactorily described using the mixed
event method (the dashed curves). The experimental distributions
were fitted by a sum of a Gaussian (with the width quoted in
Table) and the mixed background distribution. The resulting
yields, corrected for the applied cuts and the contamination from
the background $\gamma\gamma$ combinations (see \cite {ref13} for details),
are quoted in Table. \\
A rough estimation for the yield of the channel c) was extracted
from the excess of events above the combinatorial background in
the $D^+_s$ mass region (Figure 3 c). The combinatorial background
was obtained with the mixed event method and normalized to the
experimental distribution above 2.15 GeV$/c^2$ (dashed curve). The
mean yield, obtained after introducing corrections related to the
$\pi^0$ reconstruction, is quoted in Table. \\ To select
events-candidates to channel d) $\rho^+ \eta{'} (\eta{'}
\rightarrow \rho^0 \gamma)$, the following cuts were applied: the
effective mass of $\pi^+ \pi^-$ from the $\rho^0$ decay is
enclosed in the range 0.6$-$0.95 GeV$/c^2$; the effective mass of
$\pi^+ \pi^0$ from the $\rho^+$ decay is enclosed in a slightly
wider (as compared to $\rho^0$) range of 0.57$-$0.97 GeV$/c^2$, in
view of worse mass resolution ($\sigma_{\rho^+} = 47$ MeV) as
compared to $\sigma_{\rho^0} = 20$ MeV (see \cite{ref1} for
details); the effective mass of $\rho^0 \gamma$ from the $\eta{'}$
decay is enclosed in the range of $m_{\eta^{'}} \pm 2 \sigma
(\eta{'} \rightarrow \rho^0 \gamma)$, with the mass resolution
$\sigma (\eta{'} \rightarrow \rho^0 \gamma) = 30$ MeV. The
effective mass distribution for channel d) plotted in Figure 3 d
is fitted as a sum of a Gaussian (with the width quoted in Table)
and the mixed background distribution. The resulting yield,
corrected for the applied cuts, is quoted in Table. \\ To select
events-candidates to channels e) $\rho^+ \eta (\eta\rightarrow
\pi^+ \pi^- \pi^0)$ and f) $\rho^+ \eta (\eta\rightarrow
\gamma\gamma)$ the following cuts were applied: the effective
masses of $\pi^+ \pi^- \pi^0$ and $\gamma\gamma$ from the $\eta$
decay are enclosed in the ranges of $m_{\eta} \pm 2 \sigma (\eta
\rightarrow \pi^+ \pi^- \pi^0)$ and $m_{\eta} \pm 2 \sigma (\eta
\rightarrow \gamma\gamma)$, with the mass resolutions,
respectively, $\sigma (\eta \rightarrow \pi^+ \pi^- \pi^0)$ = 28
MeV \cite{ref1} and $\sigma (\eta \rightarrow \gamma\gamma)$ = 62
MeV. After these cuts, only one event-candidate to channel e) and
two event-candidates to channel f) survived. All three events had
effective masses compatible with the $D^+_s$ mass within $\pm$ 1.5
mass resolution for the corresponding channel (see Table). The
corresponding yields, corrected for the applied cuts are quoted in
Table. \\ Finally, we looked for decay channels with a $K^0_s$ in
the final state: $K^+ K^0_s$, $K^0_s K^- \pi^+ \pi^-$ and $K^+
K^0_s \pi^+ \pi^-$, with the summary branching ratio (4.09$\pm
0.22)$ \% \cite{ref2}. No combination with $z > 0.85$ was found in
the $D^+_s$ mass region, probably due to smallness of their
branching ratios and the restricted statistics of $K^0_s$ mesons
in our experiment. \\ The summary experimental yield of channels
a)$-$f) (with the summary branching ratio (27.7$\pm 3.4)$ \%) is
equal to $(1.84\pm0.48) \cdot 10^{-2}$, resulting to the total
yield of leading $D^+_s$ mesons equal to $\langle n_{D^+_s}(z >
0.85); W > 2.9$ GeV $\rangle = (6.64 \pm 1.91) \cdot 10^{-2}$.
This value can be compared with the mean yield of the vector
$D^{*-}_s$ in $\bar{\nu} Ne$- interactions measured at slightly
different kinematics: $\langle n_{D^{*-}_s}(z > 0.85); W > 3$ GeV
$\rangle = (5.1 \pm 1.6) \cdot 10^{-2}$ \cite{ref12}.

\section{Discussion and Summary}

\noindent The inclusive production of $\phi$ mesons  in
neutrinonuclear interactions is studied for the first time. It is
found that the $\phi$ production occurs predominantly in the
forward hemisphere of the hadronic c.m.s., with the mean yield
varying from $\langle n_{\phi}(x_F
> 0)\rangle = (0.92\pm0.34)\cdot 10^{-2}$ at $W > 2$ GeV up to
$(1.23\pm0.53)\cdot 10^{-2}$ at $W > 2.6$ GeV and
$(1.44\pm0.69)\cdot 10^{-2}$ at $W > 2.9$ GeV. The measured yields
are much larger than expected for the case of the direct $\phi$
neutrinoproduction, hence indicating on a dominant role of
indirect mechanisms. \\ A possible candidate to the latter are the
secondary intranuclear interactions like
\begin{equation}
M + N \rightarrow \phi + X \, ,
\end{equation}
\noindent where $M$ denotes pions, kaons, non-strange $(\rho,
\omega)$ and strange $(K^*(892))$ vector mesons, as well as $\eta$
and $\eta{'}$ mesons. To estimate the multiplicity gain $\langle
n(M \rightarrow \phi) \rangle$ of $\phi$ meson in reactions (5),
we used a simple model described and applied in
\cite{ref10,ref11,ref14}. The main quantities determining $\langle
n(M \rightarrow \phi) \rangle$ are the mean multiplicities
$\langle n_M(x_F >0) \rangle$ of intermediate mesons produced in
$\nu N$- interactions at $x_F > 0$ and the cross sections $\sigma
(M \rightarrow \phi)$ of reactions (5). \\
The values of $\langle n_{\pi}(x_F >0) \rangle$ were taken from
\cite{ref10}, subtracting the contribution from the decays of
mesonic resonances \cite{ref1}. The values of $\sigma (\pi
\rightarrow \phi)$ were taken from \cite{ref15} and averaged over
the pion momentum range from $p_{\pi} \sim 2$ GeV$/c$ up to $\sim
12$ GeV$/c$ (above which the pion yield is negligible in this
experiment). The mean value of $\sigma (\pi \rightarrow \phi)$
averaged over the pion and target nucleon species is in the range
of $\bar{\sigma}(\pi \rightarrow \phi) = 0.035\pm0.015$ mb,
resulting in $\langle n(\pi \rightarrow \phi) \rangle = (0.05\pm
0.03) \cdot 10^{-2}$. The cross sections $\bar{\sigma}(\rho
\rightarrow \phi)$ and $\bar{\sigma}(\omega \rightarrow \phi)$ are
expected to be slightly larger as compared to $\bar{\sigma}(\pi
\rightarrow \phi)$ due to larger mean momenta of $\rho$ and
$\omega$. Nevertheless, due to the smallness of their mean
multiplicities \cite{ref1}, the contribution to the $\phi$
production is approximately by one order of the magnitude smaller
as compared to $\langle n(\pi \rightarrow \phi) \rangle$. \\ The
averaged cross section for kaon-induced reactions, estimated from
the data compiled in \cite{ref16}, is in the range of
$\bar{\sigma}(K \rightarrow \phi) = 0.12\pm 0.04$ mb. The mean
multiplicity of charged kaons is assumed to be equal to that of
neutral ones taken from \cite{ref14}. The mean multiplicities of
vector mesons $K^*(890)$ were taken from \cite{ref1}, while the
cross section $\bar{\sigma}(K^* \rightarrow \phi)$ was assumed to
be equal to $\bar{\sigma}(K \rightarrow \phi)$. As a result, the
summary contribution $\langle n(K \rightarrow \phi) \rangle +
\langle n(K^* \rightarrow \phi) \rangle$ was estimated to be
$(0.01\pm0.01) \cdot 10^{-2}$. \\ The mean multiplicities of
$\eta$ and $\eta{'}$ were taken from \cite{ref1}. In view of the
$s \bar{s}$ content of $\eta$ and $\eta{'}$ (with probabilities
2/3 and 1/3, respectively), the $\phi$ production cross sections
were assumed to be related to that of $\rho^0$ production by
pions: $\bar{\sigma}(\eta \rightarrow \phi) \approx 2/3
\bar{\sigma}(\pi \rightarrow \rho^0) \approx 2$ mb and
$\bar{\sigma}(\eta{'} \rightarrow \phi) \approx 1/3
\bar{\sigma}(\pi \rightarrow \rho^0) \approx 1$ mb (at $p_{\pi}$
around a few GeV$/c$). Under this assumption, one obtains $\langle
n(\eta \rightarrow \phi) \rangle = (0.102\pm 0.054)\cdot 10^{-2}$
and $\langle n(\eta{'} \rightarrow \phi) \rangle = (0.016 \pm
0.010) \cdot 10^{-2}$, where the uncertainty in cross sections is
not included in the quoted errors. The summary contribution from
all aforesaid processes does not exceed $0.2 \cdot 10^{-2}$. This
value, even when assuming that all produced $\phi$ mesons acquire
$x_F >$ 0 (which is not the case), composes only a minor fraction
of the measured yield $\langle n_ {\phi}(x_F > 0) \rangle =
(0.92\pm 0.34)\cdot 10^{-2}$ at $W > 2$ GeV. An appreciable
contribution to the latter can be provided if only the unknown
cross sections $\bar{\sigma}(\eta \rightarrow \phi)$ and
$\bar{\sigma}(\eta{'} \rightarrow \phi)$ were a few times larger
than supposed above. It might be worthwhile to note, that an
information about the role of secondary intranuclear interactions
can be, in principle, inferred from the data (not yet existing) on
the $A$- dependence of the $\phi$ yield in
 neutrinonuclear reactions.
\\
Another source of indirect $\phi$ production at $x_F > 0$ are the
decays of leading $D^+_s$ mesons carrying the overwhelming
fraction $z$ of the hadronic energy. To look the
neutrinoproduction of $D^+_s$ mesons, six different final states
of $D^+_s$ decays are considered requiring $z > 0.85$. For all of
them, an excess of events is observed at the $D^+_s$ mass region,
allowing to infer, for the first time in neutrino-induced
reactions, an estimation for its yield: $\langle D^+_s(z > 0.85);
W > 2.9$ GeV $\rangle = (6.64 \pm 1.91) \cdot 10^{-2}$. \\
The events-candidates of the $D^+_s$ production were collected to
a subsample used furthermore for simulation of the $D^+_s
\rightarrow \phi + \rho^+$ and $D^+_s \rightarrow \phi + \pi^+$
decays composing the overwhelming fraction (85\%) of the $D^+_s
\rightarrow \phi X$ decays \cite{ref2}. The simulated differential
spectrum $dn_\phi^{dec}/dx_F$ turned out to be strongly shifted
towards the forward hemisphere in the hadronic c.m.s., providing
85.4\% of $\phi$ mesons to have $x_F > 0$ and resulting in the
expected mean yield of decay $\phi$ mesons equal to $\langle
n_F^{dec}(x_F > 0); W > 2.9$ GeV $\rangle = (0.91 \pm 0.26) \cdot
10^{-2}$. This composes more than (or about) the half of~ the~
measured~ value~ $\langle n_{\phi}(x_F > 0);W>2.9$ GeV $\rangle =
(1.44 \pm 0.69) \cdot 10^{-2}$. \\
In order to make a comparison with the measured differential
spectrum at $W > 2$ GeV (plotted in Figure 2), the yield of
$D^+_s$ was properly renormalized to the number of events with $W
> 2$ GeV, assuming that the $D^+_s$ production at $W < 2.9$ GeV
can be neglected. This lead to $\langle n_{D^+_s}(z > 0.85); W
> 2$ GeV $\rangle = (2.74 \pm 0.79) \cdot 10^{-2}$, resulting in
the expected yields of the decay $\phi$ mesons $\langle
n_{\phi}^{dec}$(all $x_F);W > 2$ GeV $\rangle = (0.44 \pm 0.13)
\cdot 10^{-2}$ and $\langle n_{\phi}^{dec}(x_F > 0);W > 2$ GeV
$\rangle = (0.38 \pm 0.11) \cdot 10^{-2}$. These compose less than
(or about) the half of experimentally estimated values,
respectively, $\langle n_{\phi}$(all~ $ x_F); W > 2$ GeV $\rangle
= (1.19 \pm 0.61) \cdot 10^{-2}$ and $\langle n_{\phi}(x_F > 0); W
> 2$ GeV $\rangle = (0.92 \pm 0.34) \cdot 10^{-2}$. As a result,
the magnitude of the predicted spectrum $dn_\phi^{dec}/dx_F$
approximately twice underestimates that for the measured one
(Figure 2). The shape of the predicted spectrum is, however,
compatible with the experimental one.

{\bf Acknowledgement.} The authors from YerPhI acknowledge the
supporting grants of Calouste Gulbenkian Foundation and Swiss
Fonds "Kidagan". The activity of one of the authors (H.G.) is
supported by Cooperation Agreement between DESY and YerPhI signed
on December 6, 2002.


\newpage
\begin{figure}[ht]
\centering
 \resizebox{0.8\textwidth}{!}{\includegraphics*[bb =90 20 490 560]
{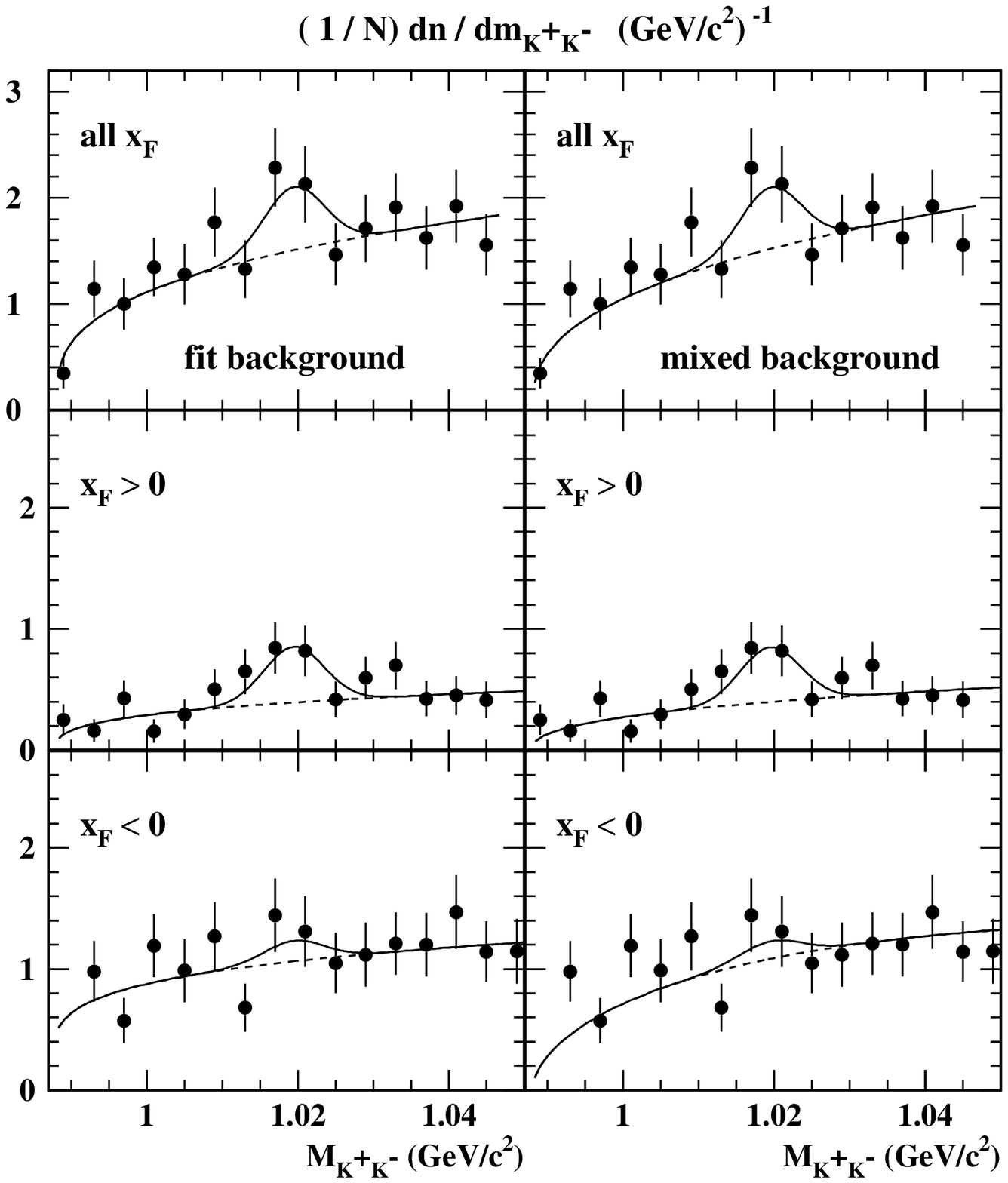}} \caption{ The $K^+K^-$ effective mass distributions at
$W > 2$ GeV. The curves are the fit results for two cases of the
background description: using the analytical form (1) (the left
panel) and the mixed event method (the right panel).}
\end{figure}

\newpage
\begin{figure}[ht]
\centering \resizebox{1.0\textwidth}{!}{\includegraphics*[bb=2 30
650 590] {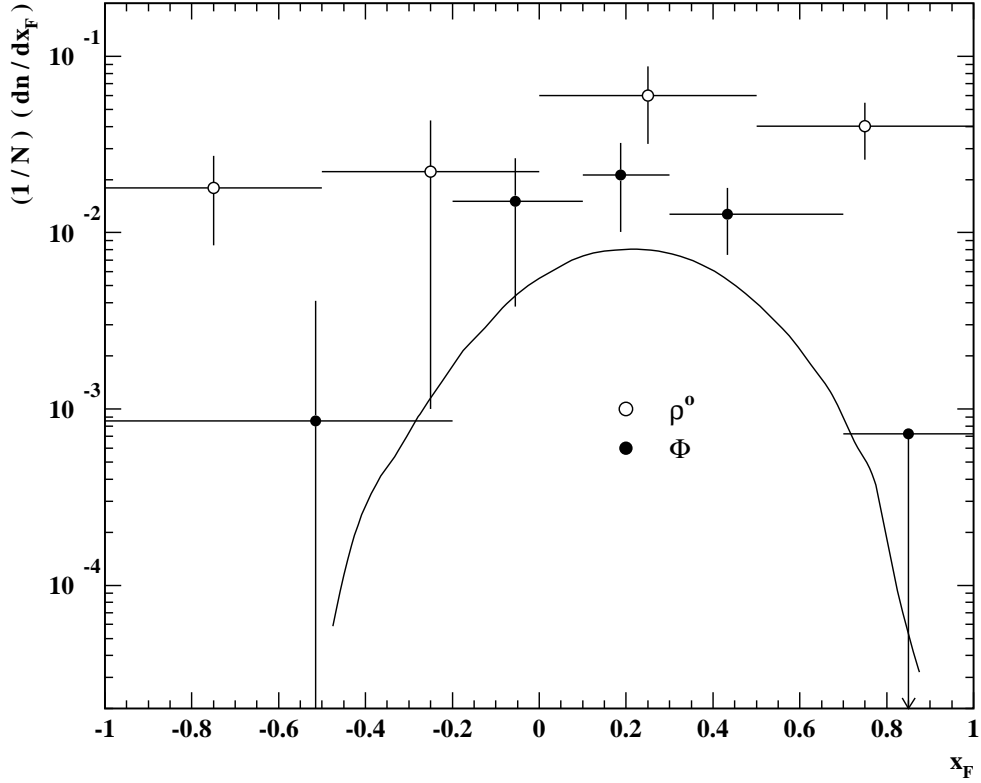}} \caption{ The $x_F$ spectra for $\phi$ and
$\rho^0$ at $W >2$ GeV. The curve is the expected distribution for
$\phi$ mesons from the $D^+_s$ decays.}
\end{figure}

\newpage
\begin{figure}[ht]
\centering \resizebox{1.0 \textwidth}{!}{\includegraphics*[bb=2 70
700 590]{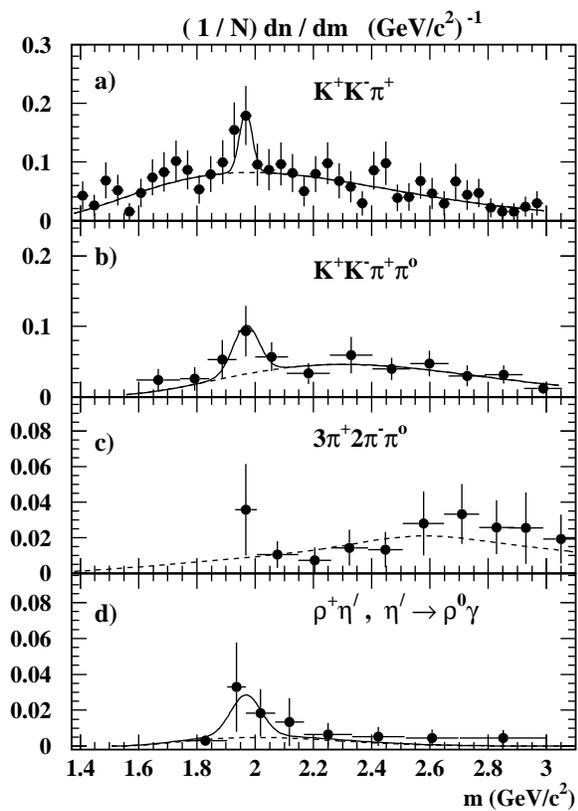}} \caption{ The effective mass distribution for
channels a)$-$d) at $W > 2.9$ GeV. The curves are the fit results
(see text).}
\end{figure}

\end{document}